\documentclass[amsmath,amssymb,a4paper,nofootinbib]{revtex4}

\usepackage{color}
\usepackage{graphicx}% Include figure files
\usepackage{dcolumn}% Align table columns on decimal point
\usepackage{bm}% bold math
\usepackage{latexsym}
\usepackage{epsfig}
\usepackage{rotating}
\usepackage{bm}% bold math

\newcommand{\be}{\begin{equation}}
\newcommand{\ee}{\end{equation}}
\newcommand{\beqq}{\setlength\arraycolsep{2pt}\begin{eqnarray}}
\newcommand{\eeqq}{\vspace{0cm} \end{eqnarray}}
\newcommand{\bea}{\begin{eqnarray}}
\newcommand{\eea}{\end{eqnarray}}

\begin{document}

\title{Cosmic transparency and acceleration }

\author{R. F. L. Holanda$^{1,2,3}$} \email{holandarfl@gmail.com}
\author{S. H. Pereira$^{4}$} \email{shpereira@feg.unesp.br}
\author{Deepak Jain$^{5}$}\email{djain@ddu.du.ac.in}
\affiliation{ \\$^1$Departamento de F\'{\i}sica, Universidade Federal de Sergipe, 49100-000 -- S\~ao Cristov\~ao - SE, Brazil,
\\$^2$Departamento de F\'{\i}sica, Universidade Federal de Campina Grande, 58429-900 -- Campina Grande - PB, Brazil,
\\$^3$Departamento de F\'{\i}sica, Universidade Federal do Rio Grande do Norte, 59078-970 -- Natal - RN, Brazil,
\\$^4$Universidade Estadual Paulista (Unesp)\\Faculdade de Engenharia, Guaratinguet\'a \\ Departamento de F\'isica e Qu\'imica\\ 
12516-410 -- Guaratinguet\'a - SP, Brazil
\\$^5$Deen Dayal Upadhyaya College, University of Delhi, Sector 3,\\ Dwarka, New Delhi 110078, India
}

%\keywords{Cosmology, Models beyond the standard model}

%\bigskip
\begin{abstract}
In this paper, by considering an absorption probability  independent of photon wavelength, we show that current type Ia supernovae (SNe Ia) and gamma ray burst (GRBs)  observations plus high-redshift  measurements of the cosmic microwave background (CMB) radiation temperature support cosmic acceleration regardless of the transparent-universe assumption. Two flat scenarios are considered in our analyses: the $\Lambda$CDM model and a kinematic model.  We consider $\tau(z)=2\ln(1+z)^{\varepsilon}$, where $\tau(z)$ denotes the opacity  between an observer at $z=0$ and a source at $z$. This choice is equivalent to deforming the cosmic distance duality relation  as $D_LD^{-1}_A = (1 + z)^{2+\varepsilon}$ and, if the absorption probability is  independent of photon wavelength, the CMB temperature evolution law is $T_{CMB}(z)=T_0(1+z)^{1+2\varepsilon/3 }$. By marginalizing on the $\varepsilon$ parameter, our analyses rule out a decelerating universe  at 99.99 \% c.l. for all scenarios considered. Interestingly, by considering only SNe Ia and GRBs observations, we obtain that a decelerated universe indicated by $\Omega_{\Lambda} \leq 0.33$ and $q_0 > 0$ is ruled out around 1.5$\sigma$ c.l. and  2$\sigma$ c.l., respectively, regardless of the transparent-universe assumption. 

\end{abstract}
\maketitle

%%%%%%%%%%%%%%%%%%%%%%%%%%%%%%%%%%%%%%%%%%%%%%%%%%%%%%%%%%%%%%%%%%%%%%%%%%

\section{Introduction}

The present stage of cosmic acceleration was proposed almost 20 years ago due to an unexpected dimming in the observed light of type Ia supernovae (SNe Ia) \cite{riess,perl}. In general relativity with a homogeneous and isotropic space-time, such cosmic behavior  requires the existence of an extra component called dark energy whose main characteristic is a negative pressure that overcomes the attractive character of matter. Nowadays, the existence of cosmic acceleration has been confirmed by several other complementary independent probes, such as the cosmic microwave background (CMB), the Hubble parameter and baryon acoustic oscillations (BAO) \cite{blake,wiggle,Ade}. However, even after 20 years we still do not know if the energy density of this unknown component is constant or varies in time and space (see the excellent reviews in Refs. \cite{weinberg,cadwell}). On the other hand, the cosmic acceleration scenario also originated the discussion about possible modifications of general relativity in order to explain the acceleration without dark energy \cite{domenico,edmund,felice}. Knowing whether dark energy exists and characterizing its equation of state is perhaps the greatest challenge in modern cosmology.

Alternative scenarios that could contribute to the evidence for this acceleration  or even mimic the behavior of dark energy have been proposed over the years. Some examples include a possible intrinsic evolution in type Ia supernovae luminosity \cite{drell,combes}, local Hubble bubble effects \cite{zehavi,conley}, or a high redshift and replenishing dust mechanism \cite{golbar,riess2004,basset}. However, these hypotheses were not supported by data (see, for instance, Sec. 4.2 in Ref. \cite{riess2004} and Refs.\cite{basset} and \cite{moss}). On the other hand, there are some cosmic opacity sources that could  influence astronomical photometric measurements. For SNe Ia observations the inferred opacities  of the following sources are model dependent: the Milky Way, the host galaxy, intervening galaxies, and the intergalactic medium \cite{combes,conley,menard,imara,McKinnon,goobar}. Another possible source is the  oscillation of photons propagating in extragalactic magnetic fields, which could convert  the photons  into very light axions  or chameleon  fields \cite{j,burrage,avg2009,avg2010,jac2010}. Although such kinds of effects are very small, they may become significant at cosmological scales, imprinting signatures on electromagnetic radiation. The contribution of these effects have been investigated in CMB, radio  and optical sources (see \cite{tiw} and references therein). In this context, the authors of Ref. \cite{lima2011} considered only SNe Ia data (Union2) and two different scenarios with cosmic absorption (epoch-dependent and -independent absorption). The authors were able to validate the $\Lambda$CDM (flat) description with a high value for the cosmological constant $\Omega_{\Lambda}$ in the extreme limit of perfect cosmic transparency (negligible cosmic absorption).

 The question of whether some cosmic opacity has influence on photometric measurements of distant SNe Ia remains open \cite{goobar}. If some extra dimming is still present, the SNe Ia observations will give us questionable values for main cosmological parameters and, consequently, for the acceleration rate. From a theoretical point of view, the opacity affects the fundamental relation between the luminosity distance $D_L$ and angular diameter distance $D_A$ the  so-called cosmic distance duality relation $D_LD_A^{-1}=(1+z)^2$ \cite{et}. In this way, several cosmic opacity tests  have been performed recently. The authors of Refs. \cite{more,deepak}  used angular diameter distances from BAO measurements and luminosity distances from SNe Ia data in a $\Lambda$CDM framework to constrain the optical depth of the Universe. The results indicated a transparent Universe, although not with significant precision. Several works exploring cosmic opacity considered a cosmic distance duality relation deformed by a $\varepsilon$ parameter, as $D_LD_A^{-1}=(1+z)^{2+\varepsilon}$. For instance,  the authors of Refs. \cite{avg2009,avg2010} used current measurements of the expansion rate $H(z)$ and SNe Ia data in a flat $\Lambda$CDM model and showed that a transparent universe is in agreement with the data considered ($\varepsilon \approx 0$). $H(z)$ data and luminosity distances of gamma ray bursts (GRBs) and SNe Ia in  $\Lambda$CDM and $\omega$CDM flat models were considered to explore the possible existence of an opacity at higher redshifts $(z > 2)$ \cite{HB2014}. Again, the results indicated the transparency of the universe, but with large error bars. Cosmological model independent analyses were also performed: the authors of Refs. \cite{HCA2013,LAZ2015}  used current measurements of the expansion rate $H(z)$ and SNe Ia data, those of Ref. \cite{li}  considered angular diameter distances from galaxy clusters and SNe Ia data, and, finally, the authors of Ref. \cite{FHA}  used 32 old passive galaxies and SNe Ia data. No significant opacity was found from these studies, although the results do not completely rule out the presence of some dimming source. 

{ For the CMB temperature evolution law,  if the cosmic expansion is adiabatic, the universe is isotropic and homogeneous, and the CMB spectrum  at $z\approx 1100$ (decoupling redshift) is that of a blackbody; it will remain a blackbody, obeying the temperature evolution law $T_{CMB}(z)=T_0(1+z)$, where $T_0$ is $2.725 \pm 0.0013$ K \cite{cobe,fixer,fixer2009}. However, although the present CMB spectrum is consistent with a blackbody (deviations are less than 50 parts per million), $T_{CMB}(z)$ measurements at intermediate and high redshifts are required to test the temperature law in the past.} This $T_{CMB}(z)$ prediction  can be violated if some mechanism acts upon this radiative component \cite{over,lima2000}. Adiabatic photon production (or destruction) or deviations from isotropy and homogeneity
could modify this scaling \cite{j}. Moreover, a source of dimming material (unless in perfect thermal equilibrium with the CMB) would tend to change its blackbody spectrum. Infrared emission can also occur after absorption of visible photons by a diffuse component of intergalactic dust, besides discrete sources (dusty star-forming galaxies) \cite{aguirre,hauser,mark}.  

Other previously cited mechanisms, such as axion-photon conversion induced by intergalactic magnetic fields \cite{mirizzi,jac2010,tiw}, could cause excessive spectral distortion of the cosmic microwave background, also, those with scalar fields with a nonminimal coupling  to the electromagnetic Lagrangian \cite{larena,r2016,r2017,r20172,r20173}, which could produce deviation from standard results. Thus, we may consider a temperature law deformed as $T_{CMB}(z)=T_0(1+z)^{1-\beta}$, where the (constant)  parameter $\beta$ has been limited by different data sets: for $z<1$, $T_{CMB}(z)$ can be obtained via the Sunyaev-Zel'dovich effect (SZE) \cite{Fabri,reph}, while for $z>1$, $T_{CMB}(z)$ has been measured from the analysis of quasar absorption line spectra \cite{r}. As a result, we may quote  $\beta = 0.041^{+0.038}_{-0.041}$ (1$\sigma$)  from Ref. \cite{lisa}  and, more recently, the authors of Ref. \cite{lisa2} found  $\beta = 0.022 \pm 0.018$ (1$\sigma$). {  In Ref.\cite{prl} the authors showed that if the cosmic distance duality relation is violated, the black body spectrum changes to a greybody spectrum. By using the FIRAS/COBE data, they put a limit of the order of 0.01\% on the possible deviation from the cosmic distance duality relation. However, this limit was obtained by using the radiation coming from the surface of last scattering at $z=1100$, so limits at low and intermediate redshifts also have to be considered.}

Very recently, the authors of Ref. \cite{avgjcap} proposed an interesting and simple relation between $\beta$ and $\varepsilon$ in the presence of a dimming agent. By considering an absorption probability  independent of photon wavelength (and preservation of the CMB blackbody spectrum), they found $\beta=-\frac{2}{3}\varepsilon$. By using the results on $\varepsilon$ from Refs. \cite{avg2009,avg2010} together with direct constraints on $\beta$ from Ref. \cite{note}, they found a competitive constraint on $\beta$: $\beta = 0.004 \pm 0.016$ (1$\sigma$). More recently, this method was performed with more data in Ref. \cite{avg2015} and the value $\beta = 0.0076 \pm 0.008$ (1$\sigma$) was obtained.

In this work, unlike in previous analyses, where the main aim was to constrain  the cosmic opacity by using SNe Ia (or GRBs) and other cosmic opacity free data sets, such as BAO and $H(z)$ we use three cosmic-transparency-dependent data sets, namely, SNe Ia \cite{bet}, GRBs \cite{demi} and $T_{CMB}(z)$, to constrain cosmological parameters by adding $\varepsilon$ as a free parameter in the analyses. In other words, we relax the cosmic transparency assumption in order to verify the cosmic acceleration. We consider two flat scenarios: $\Lambda$CDM and a kinematic model based on a parametrization of the deceleration parameter $q(z)$. By considering departures from standard cosmology, such as $D_LD_A^{-1}=(1+z)^{2+\varepsilon}$, we show that  combinations of SNe Ia, GRB observations, and $T_{CMB}$ confirm the cosmic acceleration at 99.99\% c.l. in both scenarios (marginalizing on $\varepsilon$ parameter). Our results are obtained by assuming an  absorption probability  independent of photon wavelength; in such a framework, the $T_{CMB}(z)$ measurements can be added in the analyses via a deformed temperature evolution law, such as $T_{CMB}(z)=T_0(1+z)^{1-\beta}$ (with $\beta=-\frac{2}{3}\varepsilon$) \cite{avgjcap}. By considering only SNe Ia and GRB observations we obtain that $\Omega_{\Lambda} \leq 0.33$(decelerated expansion) is ruled out around 1.5$\sigma$ c.l. and $q_0 > 0$ is ruled out at 2$\sigma$ c.l..

This paper is organized as follows: In  Section II we briefly describe the method. The cosmological data are presented in Section III, and the analyses and results are described in Section IV. Finally, we conclude in Section V.

\section{Basic equations}

\subsection{Cosmic opacity and luminosity distance}

In this section we explain the method used in this paper. The methodology used in this analyses is similar to the earlier work  in Refs. \cite{avg2009,avg2010}.  When cosmic opacity is taken into account the distance moduli derived from SNe Ia are systematically affected,  increasing their luminosity distances. If one considers $\tau(z)$ as the opacity  between an observer at $z=0$ and a source at $z$, the flux received by the observer  is  attenuated by a factor $e^{-\tau(z)}$. In this context, the observed luminosity distance ($D_{L, obs}$) is related to the true luminosity  distance ($D_{L, true}$) by
 \begin{equation}
 D_{L, obs}^2=D_{L,true}^2 e^{\tau(z)} \, .
 \end{equation}
Thus, the \emph{observed} magnitude distance modulus is given by 
\begin{equation}
\label{distancemod}
m_{obs}(z)=m_{true}(z)+2.5(\log e) \tau(z) \, .
\end{equation}
 
In this work we perform the analysis based on luminosity distance from two different approaches:
a flat $\Lambda$CDM cosmology and a kinematic model with a parametrization for the deceleration parameter $q(z)$.

\subsubsection{ The flat $\Lambda$CDM model } 

In a flat $\Lambda$CDM model the luminosity distance is given by
\begin{equation}
\label{dl}
 D_{L,true}(z)=(1+z)c\int_0^z\frac{dz'}{H(z')},
 \end{equation}
 where $c$ is the speed of light and
  \begin{equation}
	\label{hz}
  H(z)=H_0 E(z,\textbf{p})=H_0[\Omega_M(1+z)^3+(1-\Omega_M)]^{1/2}.
  \end{equation}
In the above expressions, $\Omega_M = 1-\Omega_\Lambda$ stands for the matter density parameter measured today, $\Omega_\Lambda$ is the cosmological constant density parameter and $H_0$ is the Hubble constant.

\subsubsection{ The kinematic approach} 

{  For the kinematic approach, we consider the deceleration parameter, given by $q=-\frac{\ddot{a}}{aH^2}$, where $H=\dot{a}/a$ is the Hubble parameter and $a(t)=(1+z)^{-1}$, from which one may write }
\begin{equation}
 q(z)=\frac{1+z}{H}\frac{dH}{dz}-1={1\over 2}\frac{d \ln H^2}{d \ln(1+z)}-1\,. \label{qz}
\end{equation}
In this approach, we use one of the most frequently used parametrization of $q(z)$ \cite{chavallier,linder,santos2010},
\begin{equation}
 q(z)=q_0+q_1\frac{z}{1+z},\label{qz1}
\end{equation}
where $q_0$ is the current deceleration parameter value and the second term has the property that in the infinite past $q\to q_0+q_1$. Such a parametrization mimics the behavior of a wide class of
 accelerating dark energy models.  Since we expect the Universe to be matter-dominated at early times, which implies $q(z\gg 1)=1/2$, we have $q_0+q_1=1/2$, from which Eq. (\ref{qz1}) becomes
\begin{equation}
 q(z)=\bigg(q_0+{z\over 2}\bigg)\frac{1}{1+z}\label{qz2}.
\end{equation}
Then, the luminosity distance in terms of $q$ can be written as
\begin{equation}
\label{dlq0}
 D_{L,true}(z)={(1+z)c\over H_0}\int_0^z \exp\bigg[-\int_0^u [1+q(u)]d\ln(1+u)\bigg]du.
 \end{equation}

\begin{figure*}[t]
\centering
\includegraphics[width=0.3\textwidth]{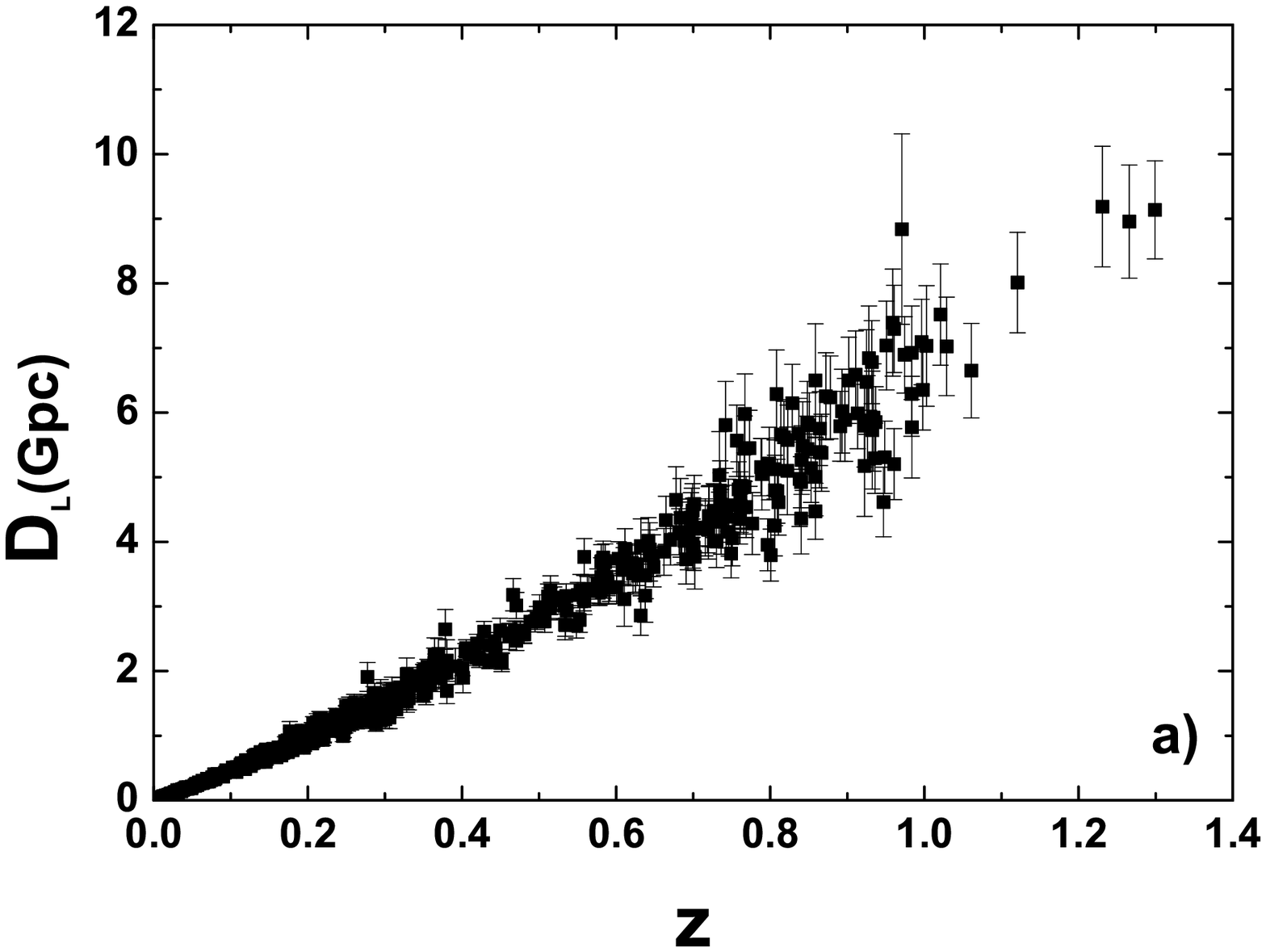}
\hspace{0.3cm}
\includegraphics[width=0.3\textwidth]{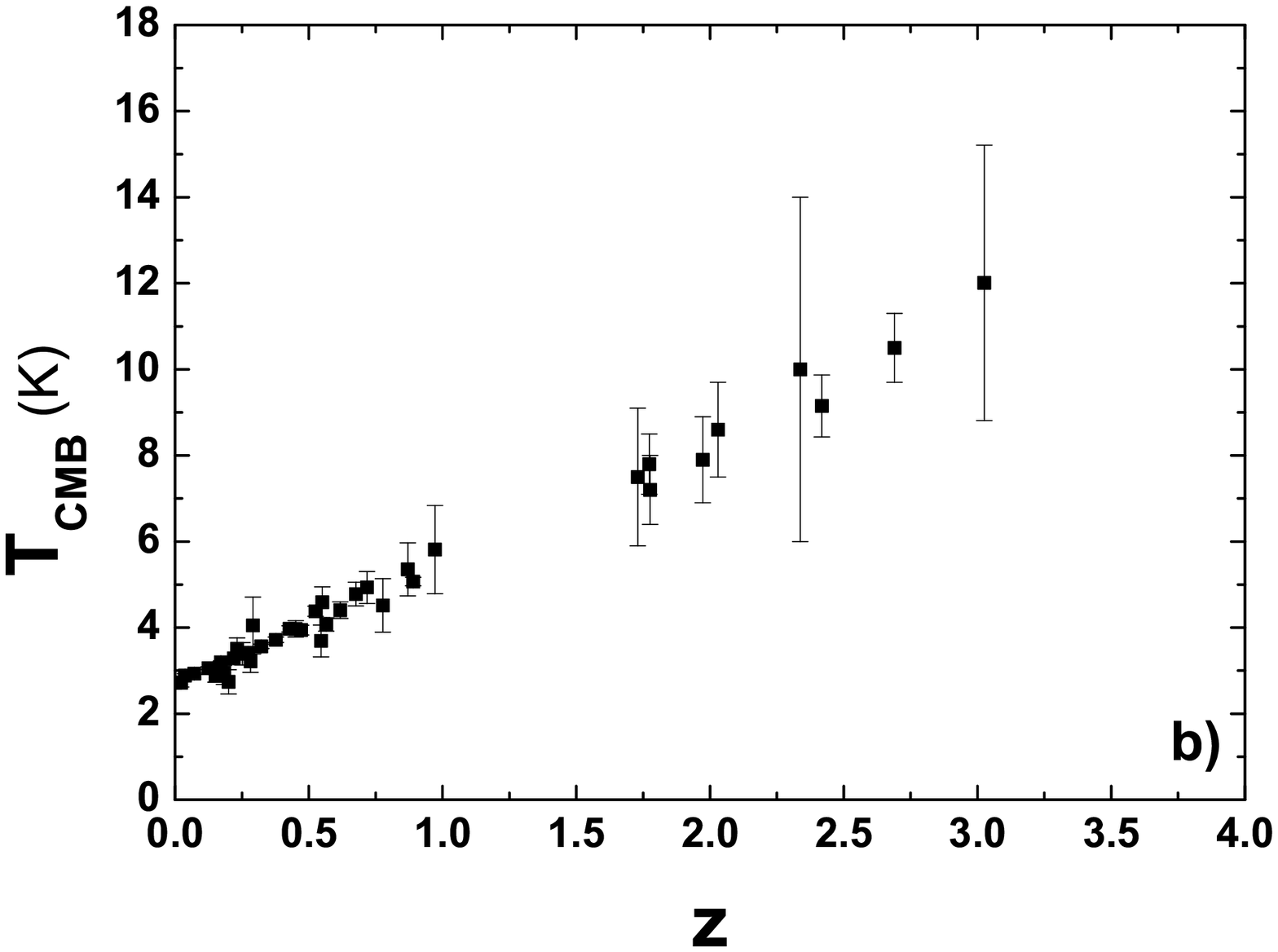}
\hspace{0.3cm}
\includegraphics[width=0.3\textwidth]{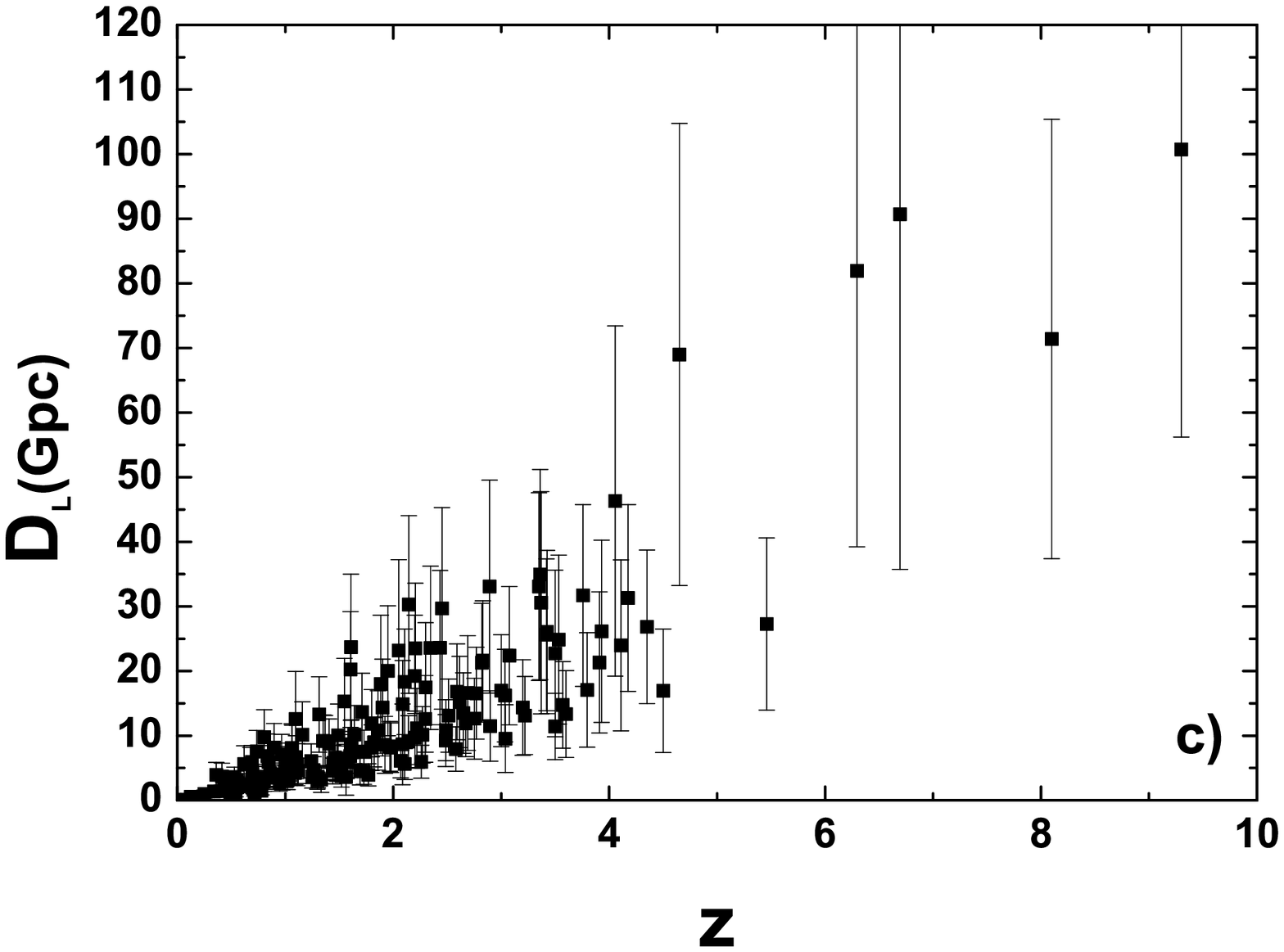}
\caption{ In Fig.(a) we plot the 740 luminosity distances  of SNe Ia \cite{bet}. In Fig.(b) we plot the 42 $T_{cmb}$ measurements. In Fig.(c) we plot the 162 luminosity distances  of GRBs \cite{demi}.}
\end{figure*}

At this point it is worth stressing that in previous work the cosmic opacity was taken as $\tau(z)=2\varepsilon z$. The authors argued that for small $\varepsilon$ and $z \leq 1$ this is equivalent to deforming the cosmic distance duality relation (CDDR) as $D_LD_A^{-1}=(1+z)^{2+\varepsilon}$. However, in order to obtain more robust results, we do not  exactly follow this approach, as we are using data sets that are also present at higher redshift. So, a complete expression for the \emph{observed} distance modulus is used, which can be obtained from the deformed CDDR, $D_{L, obs}=D_{L,true}(1+z)^{\varepsilon}$,  such as
\begin{equation}
\label{distancemod2}
m_{obs}(z)=m_{true}(z)+5\log(1+z)^{\varepsilon} \, .
\end{equation} 
Then, the $\tau(z)$ function in our case is  $\tau(z)=2\ln(1+z)^{\varepsilon}$.

\subsection{Cosmic opacity and CMB temperature law}

We have seen that luminosity distance is affected by a departure from cosmic transparency, since flux goes inversely with the square of luminosity distance. Changes also occur in the CMB temperature law. Deviations of the standard CMB temperature law  have been written  using the parametrization $T_{CMB}(z)=T_0(1+z)^{1-\beta}$, where $\beta$ is a constant. This scaling is presumed to be a consequence of photon number and radiation entropy nonconservation \cite{lima1996,lima2000}.  Actually, $\beta$ may be a  function of redshift, but since a possible deviation is expected to be small,  a constant $\beta$ can be  justified by current error bars. In this context, the authors of Ref. \cite{avgjcap} considered the CMB spectrum in presence of a dimming source (as dust or axion-photon conversion) and by assuming that the photon survival probability is independent of wavelength, they found a direct relation between $\varepsilon$ and $\beta$: $\beta=-\frac{2}{3}\varepsilon$. In this way, we have
\begin{equation}
\label{temp22}
T_{CMB}(z,\varepsilon)=T_0(1+z)^{1+2 \varepsilon/3}.
\end{equation}
This equation and Eq. (\ref{distancemod2}) are crucial for our analyses. However, as it is well known, a photon dimming is expected to be stronger at high photon energies, so one must be careful when using indirect bounds on  SNe Ia or GRB  brightness  coming from $T_{CMB}(z)$ measurements. In this way, in Table I we show recent constraints on $\varepsilon$ parameter from observations at different wavelengths: 
optical from SNe Ia observations, microwave obtained from $T_{CMB}(z)$ measurements considered in this paper (see next section) by using Eq. (\ref{temp22}), X-ray from gas mass fraction (GMF) measurements  and gamma-ray from GRBs observations. {  As one may see, the  obtained values  for $\varepsilon$ from different wavelengths are in full agreement with each other within 1$\sigma$, so, the relation $\beta=-\frac{2}{3}\varepsilon$ is  verified with observational data and it will be used in our analyses. 

At this point, it is worth commenting that the authors of Ref.\cite{Syksy} considered a source with a blackbody spectrum (BBS) at temperature $T_{BBS}$ and they obtained that if $T_{BBS}(z)=T_{0BBS}(1+z)^{1+\varepsilon}$, the CDDR for blackbody sources should change for $D_L=D_A(1+z)^{2(1+\varepsilon)}$. The authors introduced a species of dark radiation particles to which photon energy density is transferred and obtained $ \varepsilon < 4.5 \mbox{x} 10^{-3}$ at 2$\sigma$ c.l. by using Planck data \cite{Ade}. However, Eq. (\ref{temp22}) is more general since it relates possible departures from the standard CDDR (at any band in the electromagnetic spectrum used to obtain $D_L$) with possible departures from the CMB temperature evolution law. The basic assumption is that the probability that photons are created or destroyed is independent of their wavelength.}

\begin{table*}[htb]
\caption{Constraints on $\varepsilon$ from several wavelength observations. The values of works with the symbol $*$ were obtained considering an absorption probability  independent of photon wavelength.}
{\begin{tabular} {c||c||c|c|c}
Reference & Data set & Model & $\tau(z)$&$\varepsilon$ ($1\sigma$) 
 \\
\hline \hline 
\cite{avg2009}& 307 SNe Ia + 10 $H(z)$ & flat $\Lambda$CDM &  $2\varepsilon z$ &$-0.01^{+0.06}_{-0.04}$ \\
\cite{avg2010}& 307 SNe Ia + 12 $H(z)$&flat $\Lambda$CDM&  $2\varepsilon z$ &$-0.04^{+0.04}_{-0.03}$  \\
\cite{HCA2013} &    581 SNe Ia + 28 $H(z)$ & model independent&  $2\varepsilon z$&$0.017 \pm 0.052$ \\
\cite{HB2014}& 581 SNe Ia + 19 $H(z)$ & flat $\Lambda$CDM &  $\varepsilon z$&$0.02 \pm 0.055$\\
\cite{HB2014}& 59 GRB +   19 $H(z)$ & flat $\Lambda$CDM & $\varepsilon z$ &$0.06 \pm 0.18$\\
\cite{HB2014}& 581 SNe Ia +19 $H(z)$ & flat XCDM & $\varepsilon z$ &$0.015 \pm 0.060$\\
\cite{HB2014}& 59 GRB + 19 $H(z)$ &flat XCDM &  $\varepsilon z$&$0.057 \pm 0.21$ \\
\cite{LAZ2015}& 740 SNe Ia + 19 $H(z)$ & model independent & $2\varepsilon z$& $0.044^{+0.078}_{-0.080}$\\
\cite{kamila}& 40 $GMF$ + 38 $H(z)$ &flat $\Lambda$CDM &  $2\varepsilon z$&$0.03\pm 0.08$\\
\cite{kamila}& 42 $GMF$ + 38 $H(z)$ &flat $\Lambda$CDM &  $2\varepsilon z$&$0.05 \pm 0.13$\\
\cite{FHA}& 580 SNe Ia + 32 galaxy ages & model independent & $2\varepsilon z$&$0.016 \pm 0.040$\\
\cite{lisa}$^*$& 13 $T_{CMB}(z)$ & model independent & $2\ln(1+z)^{\varepsilon} $ & $-0.06^{+0.060}_{-0.60}$\\
\cite{lisa2}$^*$& 104 $T_{CMB}(z)$&  model independent& $2\ln(1+z)^{\varepsilon} $& $-0.033 \pm 0.027$  \\
This paper$^*$ & 42 $T_{CMB}(z)$   & model independent &  $2\ln(1+z)^{\varepsilon} $&$-0.02\pm 0.04$\\
\hline
\end{tabular}} \label{ta2}
\end{table*}

\section{Data}

In this paper, we consider the following data sets:

\begin{itemize}
\item 740 SNe Ia distance moduli from the JLA compilation \cite{bet} obtained through the SALT II fitter \cite{guy}. The distance modulus of a SNe Ia is obtained by a linear relation from its light curve as $\mu = m_b -(M- \alpha  x_1 + \beta  C)$, where $m_b$ is the observed rest-frame B-band peak magnitude of the SNe Ia, $x_1$ is the time stretching of the light curve, $C$ is the supernova color at maximum brightness, $M$ is the  absolute magnitude, and $\alpha$ and $\beta$ are the shape and color corrections of the light curve. One can fix the values of $M$, $\alpha$ and $\beta$ for different models. Here we use the bounds on these parameters given by \cite{bet} for $\Lambda$CDM model: $M=-19.05 \pm 0.02$, $\alpha=0.141 \pm 0.006$ and $\beta=3.101 \pm 0.075$. It has been observed that $\alpha$ and $\beta$ act like global parameters, regardless of the prior cosmological model one chooses to determine the distance modulus of each SNe Ia (see Fig. 1a). The data points include statistical plus systematic errors.

\item 162 GRB distance moduli from Ref. \cite{demi}. This sample is in redshift range $0.033 \leq z \leq 9.3$ (see Fig. 1c). Basically, the authors of Ref. \cite{demi} used the Union2.1 compilation at redshifts close to GRBs to calibrate the Amati relation \cite{amati}. It relates the peak photon energy of a GRB  with its isotropically equivalent radiated energy.  By fitting a power law with an intrinsic
scatter where SNe Ia and GRBs overlap, the parameters of the fit could be used to determine the distance moduli of the GRBs at higher redshifts
and their respective uncertainty. A possible redshift dependence of the correlation and its effect on the GRB Hubble diagram was tested and no significant redshift dependence was found.  

\item 42 $T_{CMB}(z)$ data. The current CMB temperature, $T_0 = 2.725\pm 0.0013$ K, is estimated from the COBE satellite \cite{cobe,fixer,fixer2009}. From the method based on multi-requency SZE observations towards galaxy clusters we have used  31 data: 13  from Ref. \cite{lisa} ($0.023 \leq z \leq 0.55$) and 18 from Ref. \cite{hurier} ($0.037 \leq z \leq 0.972$). For high redshifts (10 points) we have used $T_{CMB}(z)$ obtained from observations of spectral lines \cite{e1,e2,e3,e4,e5,e6}. In total, this represents 42 observations of the CMB temperature at redshifts between 0 and 3.025 (see Fig. 1b).
\end{itemize}

\section{Analyses and results}

We obtain the constraints on the set of parameters ${\mathbf{p}}$, where  $\mathbf{p}=(\Omega_{\Lambda})$ and $\mathbf{p}=(q_0)$ when the flat $\Lambda$CDM and the kinematic approach are considered, respectively, based on the evaluation of the likelihood distribution function, ${\cal{L}} \propto e^{-\chi^{2}/2}$, with

\begin{eqnarray}
\chi^{2} &=&
 \sum_{i=1}^{740}\frac{[m_{obs}(z_i) - 5\log D_{L,true}(z_i,\mathbf{p})-\mu_0-5\log(1+z_i)^{\varepsilon}]^2}{\sigma^2_{m_{SNobsi}}}
+ \sum_{i=1}^{42}\frac{[T_{obs}(z_i)-T_{CMB}(z_i,\varepsilon)]^2}{\sigma^2_{T_{obsi}}} \nonumber \\
&+& \sum_{i=1}^{162}\frac{[m_{obs}(z_i) - 5\log D_{L,true}(z_i,\mathbf{p})-\mu_0-5\log(1+z_i)^{\varepsilon}]^2}{\sigma^2_{m_{GRBobsi}}}
  \end{eqnarray}
where  $\sigma^2_{m_{SNobsi}}$, $\sigma^2_{m_{GRBobsi}}$ and $\sigma_{T_{obsi}}^2$  are the error associated to SNe Ia and GRB distance moduli and the  error of the $T_{CMB}(z)$ measurements, respectively. $D_{L,true}$ is given by equations (\ref{dl}) and (\ref{dlq0}) for $\Lambda$CDM and kinematic frameworks, respectively,  $T_{CMB}(z,\varepsilon)$ is given by Eq. (\ref{temp22}). The quantity $\mu_0$ is the so-called nuisance parameter: $\mu_0=25-5\log H_0$.

As is commonly done in the literature, all of the results in our analyses from SNe Ia and GRB data are derived by marginalizing the likelihood function over the pertinent nuisance parameter \cite{riess2004,rapeti} (see also the section III in Ref.\cite{jaslima}). Then, one may obtain a new likelihood distribution function, ${\cal{\widetilde{L}}} \propto e^{-\widetilde{\chi}^{2}/2}$, where $\widetilde{\chi}^{2}$ is given by
\begin{eqnarray}
\widetilde{\chi}^{2}  &=& a_{SNe} - \frac{b_{SNe}^2}{c_{SNe}} + \ln(\frac{c_{SNe}}{2\pi})
  + a_{GRB} - \frac{b_{GRB}^2}{c_{GRB}}+ \ln(\frac{c_{GRB}}{2\pi})
 + \sum_{i=1}^{42}\frac{[T_{obs}(z_i)-T_{CMB}(z_i,\varepsilon)]^2}{\sigma^2_{T_{obsi}}},
  \end{eqnarray}
with	
\begin{eqnarray}
a_{SNe/GRB}&=& \sum_{SNe/GRB}\frac{[m_{(SNe/GRB)obs}(z_i)- m^{*}(z_i,\mathbf{p})]^2}{\sigma^2_{m_{SNe/GRBobsi}}} \nonumber \\
 b_{SNe/GRB}& = & \sum_{SNe/GRB}\frac{[m_{(SNe/GRB)obs}(z_i)-m^{*}(z_i,\mathbf{p} )]}{\sigma^2_{m_{SNe/GRBobsi}}} \nonumber \\
c_{SNe/GRB}&=&  \sum_{SNe/GRB}\frac{1}{\sigma^2_{m_{(SNe/GRB)obsi}}}.
  \end{eqnarray}	
In this equation $m^{*}(z_i,\mathbf{p})=5\log D_{L,true}(z_i,\mathbf{p})+5\log(1+z_i)^{\varepsilon}$.

Our results for the flat $\Lambda$CDM model from the SNe Ia and GRB observations are plotted in Fig.(2a). All contours are for 68.3\%, 95.4\% and 99.73\% c.l.. By using exclusively the SNe Ia data (black solid line), we obtain $\Omega_{\Lambda}=0.58^{+ 0.37 + 0.40}_{-0.37 - 0.56}$ and the absorption parameter $\varepsilon=0.06^{+ 0.28 + 0.37}_{-0.31 - 0.37}$ for 68.3\% and 95.4\% c.l.. As one may see, although distance moduli of  SNe Ia have been obtained in a $\Lambda$CDM framework, the analysis by using only SNe Ia data  supports a decelerated expansion even within 68.3\% c.l. if some cosmic absorption mechanism is present. Moreover, the Einstein-de Sitter model ($\Omega_{M}=1$) and de Sitter model ($\Omega_{M}=0$) are allowed almost within 95.4\% c.l.. Thus, as stressed in Ref.  \cite{lima2011}, an accelerating dark energy component must be invoked via SNe Ia data only in a transparent universe. By using exclusively the GRB data (red dash-dotted line), we obtain $\Omega_{\Lambda}\leq 0.79$ for 68.3\% and no limits are obtained for 95.4\% and 99.73\% c.l. on the parameter space. The absorption parameter in this case is: $\varepsilon=0.17^{+ 0.13 + 0.22}_{-0.13 - 0.22}$ for 68.3\% and 95.4\% c.l.. From the joint analysis by using SNe Ia and GRB observations (filled region) one may see that a decelerated universe is allowed only within  95.4\% c.l.. We obtain $\Omega_{\Lambda}=0.61^{+ 0.25 + 0.35 + 0.39}_{-0.26 - 0.42 - 0.56}$ for 68.3\%, 95.4\% and 99.73\% c.l. (two free parameters).

 Now, if one assumes an absorption probability  independent of photon wavelength,  $T_{CMB}(z)$ measurements can be added to analyses via a deformed temperature evolution law, such as $T_{CMB}(z)=T_0(1+z)^{1+\frac{2}{3}\varepsilon}$ \cite{avgjcap}. In Fig.(2b), from $T_{CMB}(z)$ data (blue dashed line),  the absorption parameter is $\varepsilon=-0.02 \pm 0.04 \pm 0.07$ for 68.3\% and 95.4\% c.l. and no limit on $\Omega_{\Lambda}$ is obtained. On the other hand, in a joint analysis involving SNe Ia, GRB data, as well as $T_{CMB}(z)$, tighter limits on $(\Omega_\Lambda,\varepsilon$) plane are obtained and a cosmic acceleration is allowed with high confidence level (filled region) even if there is some cosmic achromatic absorption mechanism. In this case we obtain: $\Omega_{\Lambda}=0.74^{+ 0.04 + 0.06 + 0.11}_{-0.04 - 0.07 - 0.12}$ for 68.3\%, 95.4\% and 99.99\% (two free parameters). In Fig.(4a) we plot the likelihood of $\Omega_{\Lambda}$ by marginalizing on the $\varepsilon$ parameter (by using a flat prior $-1 \leq \varepsilon \leq 1$). As one may see,  by considering only SNe Ia and GRB observations we obtain that $\Omega_{\Lambda} \leq 0.33$(decelerated expansion) is ruled out around 1.5$\sigma$ c.l.\footnote{{  In figs. 4a and 4b the blue curves are non-Gaussian, so the horizontal black lines provide only approximate values ​​for the intervals of 1 and 2 $\sigma$ c.l..}}.

\begin{figure*}[t]
\centering
\includegraphics[width=0.47\textwidth]{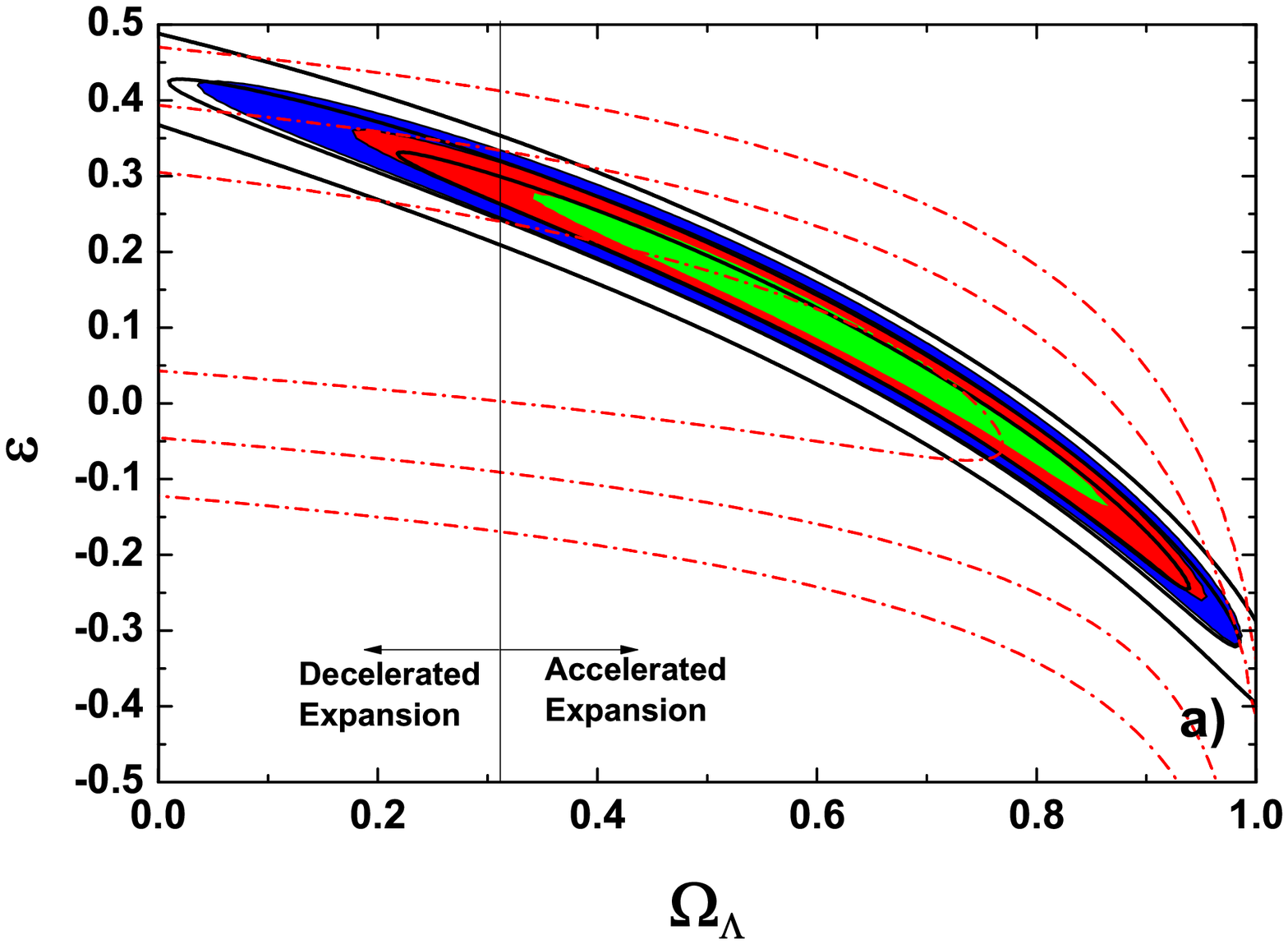}
\hspace{0.3cm}
\includegraphics[width=0.47\textwidth]{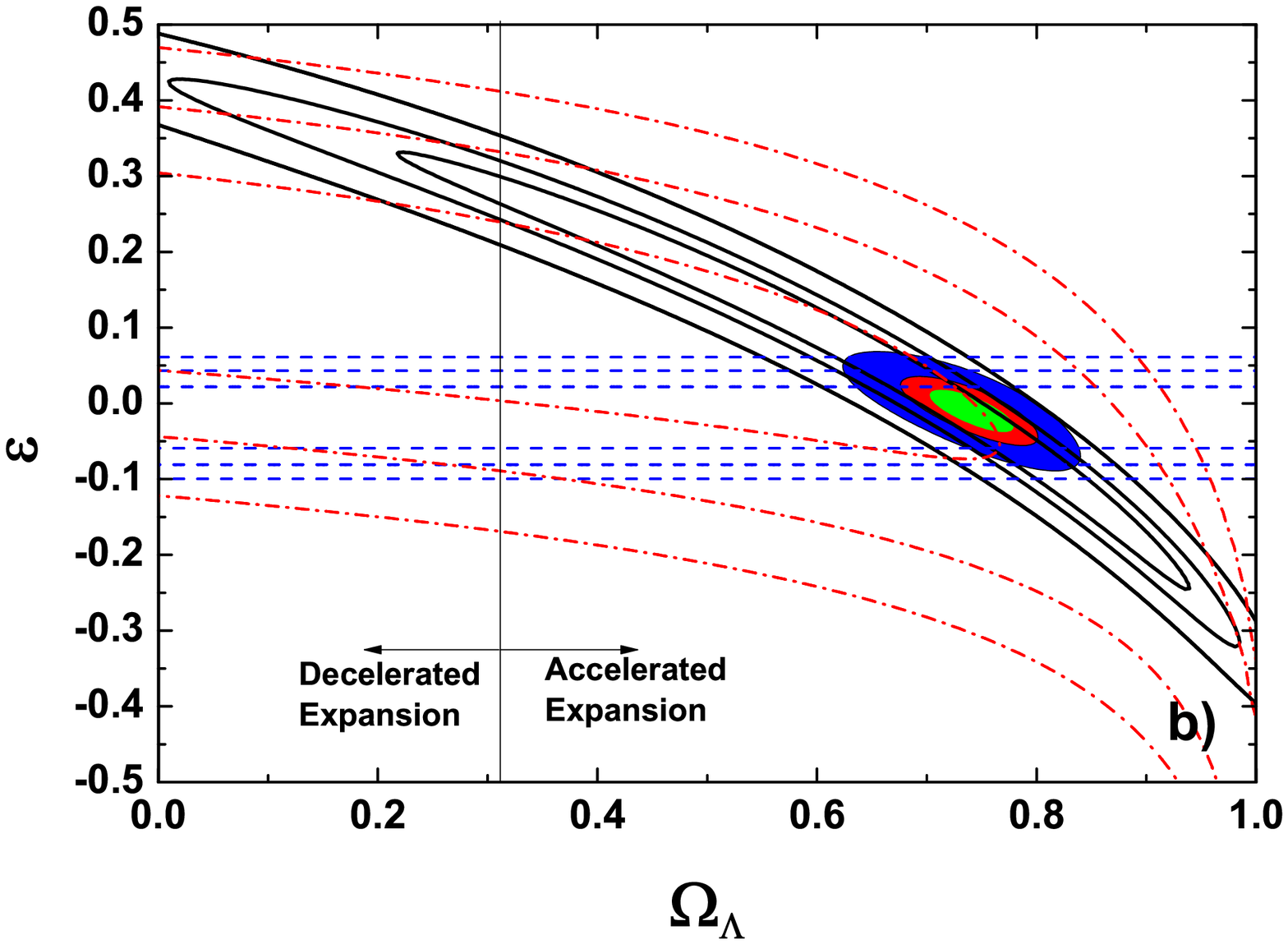}
%\hspace{0.3cm}
%\includegraphics[width=0.33\textwidth]{fig6.eps}
\caption{ In Fig.(a) the black  solid and  red dash-dotted lines correspond the analyses using SNe Ia and GRBs, respectively, within the flat $\Lambda$CDM framework. In Fig.(b) the black  solid,  red dash-dotted and blue dashed lines correspond to the analyses  using SNe Ia, GRBs and $T_{CMB}(z)$, respectively, within the flat $\Lambda$CDM framework. The filled contours are  from the joint analysis. All contours are for 68.3\%, 95.4\% and 99.73\% c.l. (except for the filled region in Fig.(b), for this case the contours are for 68.3\%, 95.4\% and 99.99\%).  }
\end{figure*}

\begin{figure*}
\centering
\includegraphics[width=0.47\textwidth]{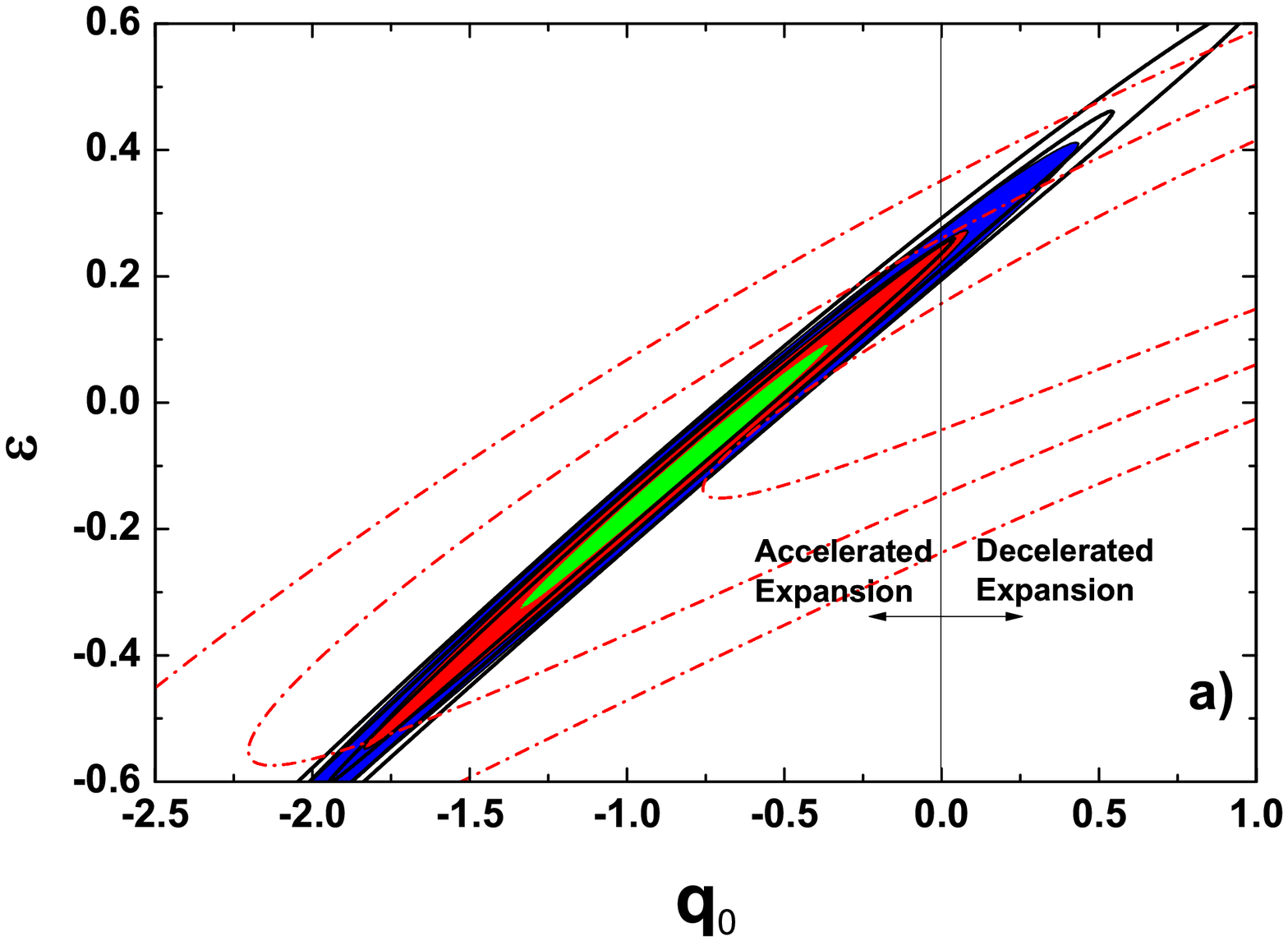}
\hspace{0.3cm}
\includegraphics[width=0.47\textwidth]{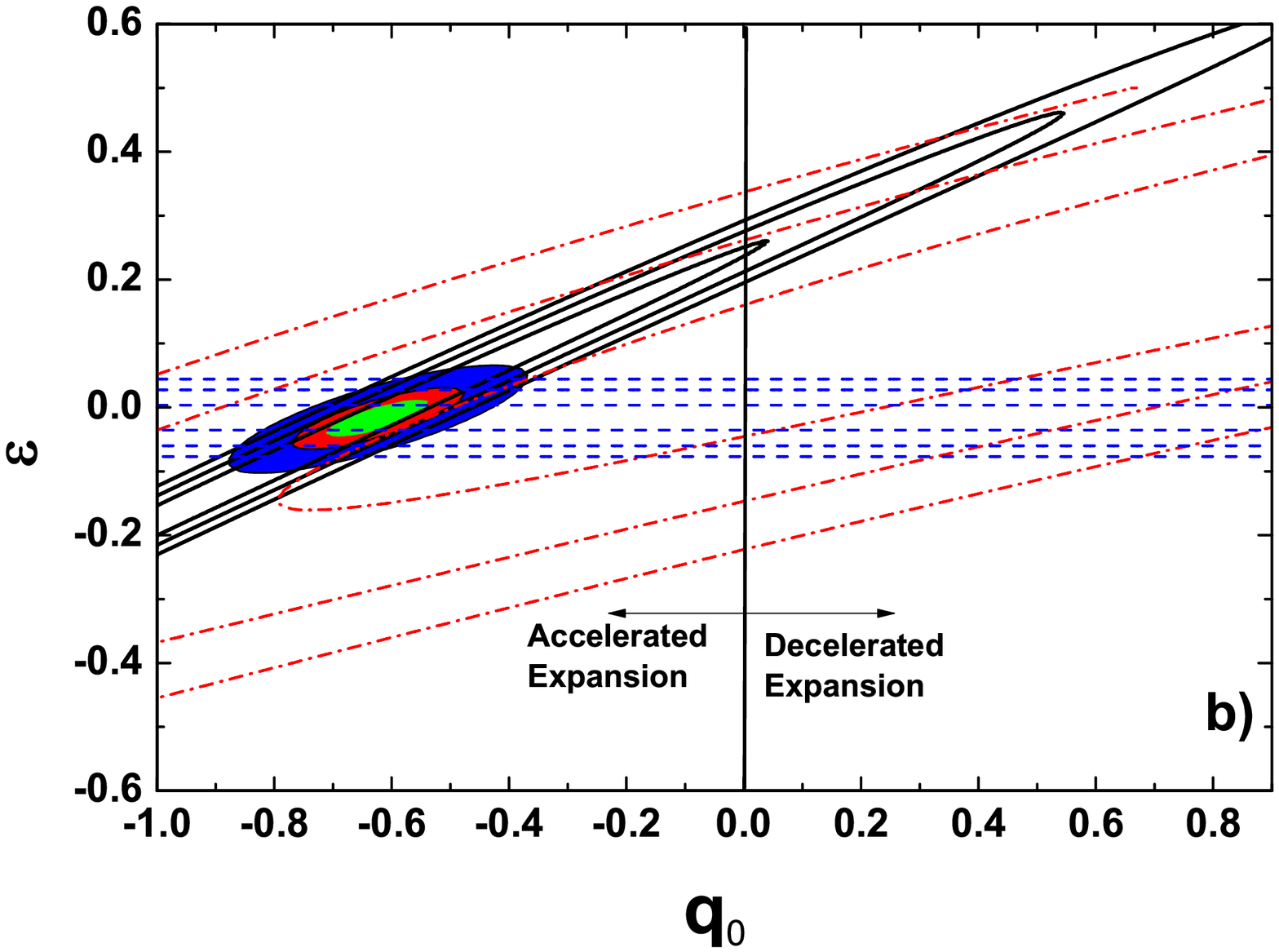}
%\hspace{0.3cm}
%\includegraphics[width=0.33\textwidth]{fig9.eps}
\caption{ In Fig.(a) the black  solid and  red dash-dotted lines correspond the analyses using, separately, SNe Ia and GRBs, respectively, within the kinematic model. The filled contours are  from the joint analysis. In Fig.(b) the black  solid,  red dash-dotted and blue dashed lines correspond the analyses by using, separately, SNe Ia, GRBs and $T_{CMB}(z)$, respectively within the kinematic framework. The filled contours are  from the joint analysis. All contours are for 68.3\%, 95.4\% and 99.73\% c.l. (except for the filled region in Fig.(b), for this case: 68.3\%, 95.4\% and 99.99\%). }
\end{figure*}

\begin{figure*}
\centering
\includegraphics[width=0.47\textwidth]{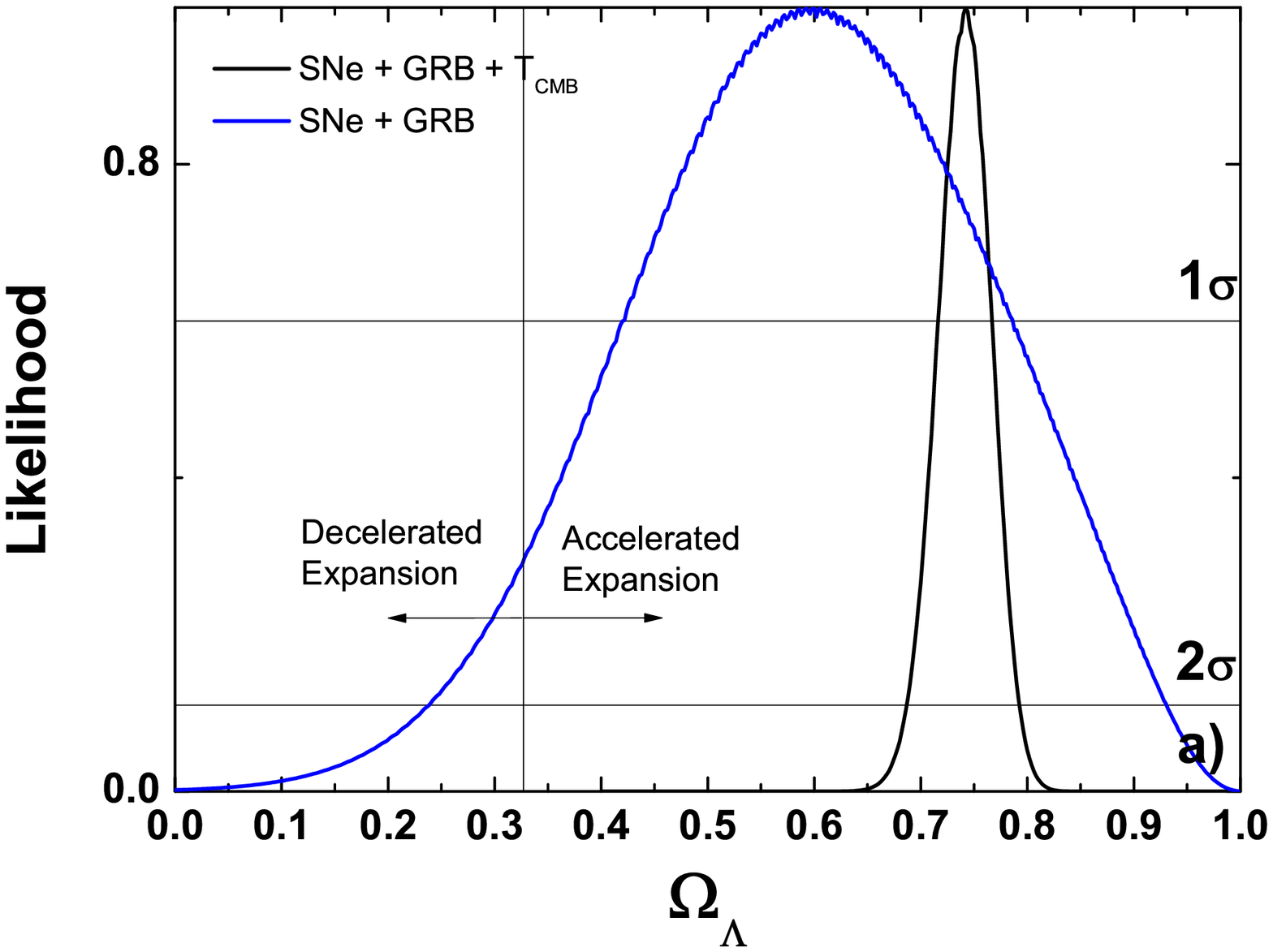}
\hspace{0.3cm}
\includegraphics[width=0.47\textwidth]{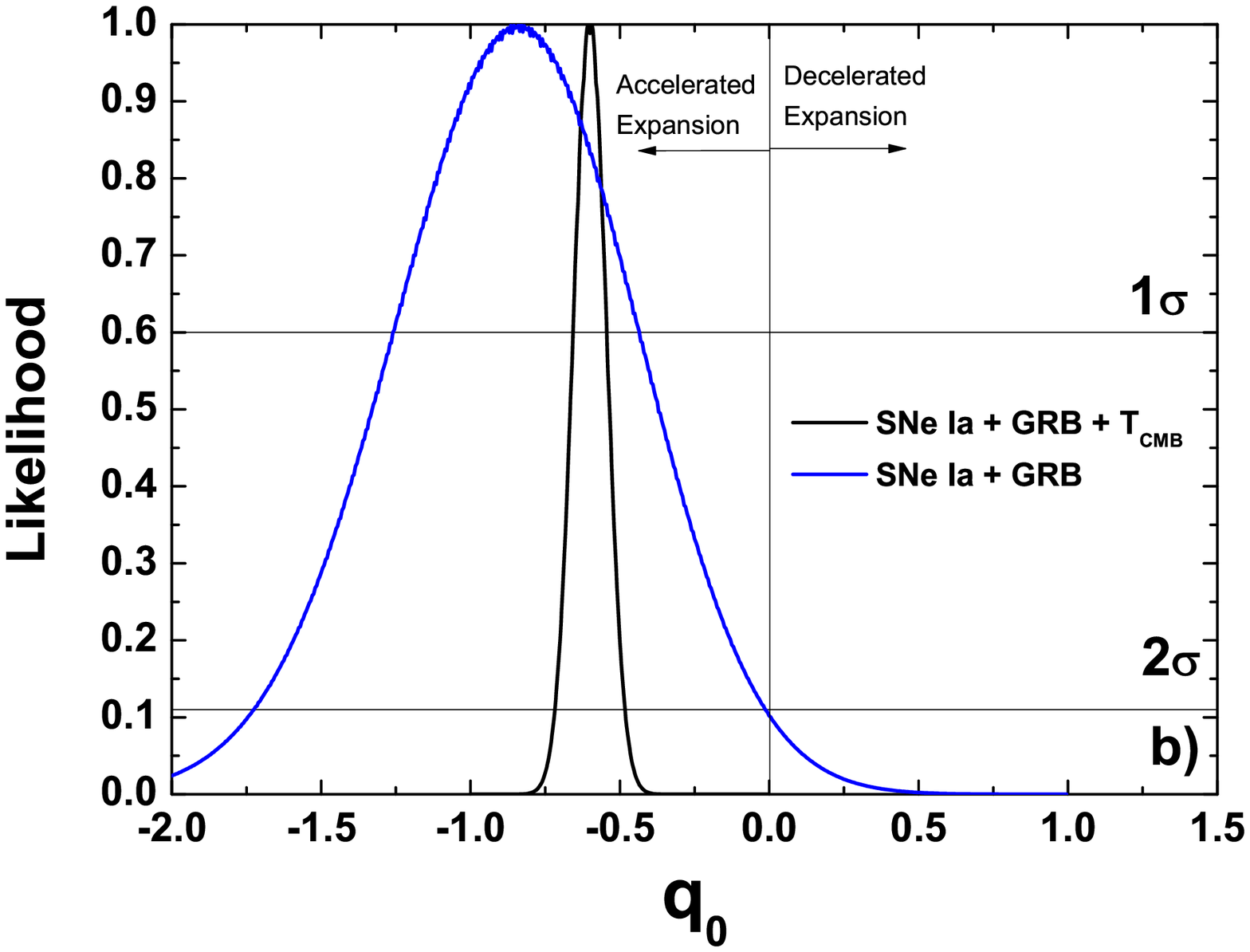}
%\hspace{0.3cm}
%\includegraphics[width=0.33\textwidth]{fig6.eps}
\caption{ In Fig.(a) we plot the likelihoods for $\Omega_{\Lambda}$ by marginalizing on $\varepsilon$  parameter by using only SNe Ia plus GRB observations (blue line) and SNe I,  GRBs plus $T_{CMB}(z)$ (black line). In Fig.(b) we plot the likelihoods for $q_{0}$ by marginalizing on $\varepsilon$  parameter by using only SNe Ia plus GRBs observations (blue line) and SNe I,  GRBs plus $T_{CMB}(z)$ (black line). The horizontal lines correspond to 1$\sigma$ c.l. (68.3\%) and 2$\sigma$ c.l. (95.4\%).}
\end{figure*}

Our results for kinematic  model are plotted in Figs.(3a) and (3b). In Fig.(3a) the nonfilled contours are for 68.3\%, 95.4\%  and 99.73\% c.l..  By using exclusively the SNe Ia data (black solid line), we obtain $q_0 \leq 0$ around 68.3\% c.l., but within 95.4\% a decelerated universe is allowed by the data. For  99.73\% no limit is obtained within the parameter space. In particular, the Einstein-de Sitter model ($q_0=1/2$) is allowed  within 95.4\% c.l.. As one may see, the SNe Ia data alone support a decelerated expansion  within  68.3\% c.l. if some cosmic absorption mechanism is present. By using exclusively the GRB data (red dash-dotted line), we obtain $q_0 \geq - 0.75$ for 68.3\% and $q_0 \geq - 2.23$ for 95.4\%. The best fit for the absorption parameter in this case is: $\varepsilon=0.18$. The joint analysis is plotted in the filled regions. Then, by using SNe Ia and GRB observations (filled region), one may see that a decelerated universe is  ruled out by the data around 95.4\% c.l.. We obtain $q_{0}=-0.80^{+ 0.40 + 0.85}_{-0.45 - 0.90}$ (two free parameters).

Again, from $T_{CMB}(z)$ data,  the absorption parameter is $\varepsilon=-0.02 \pm 0.04 \pm 0.07$ for 68.3\% and 95.4\% c.l. and no limit on $q_0$ is obtained.   On the other hand, in a joint analysis involving SNe Ia, GRB data as well as $T_{CMB}(z)$, tighter limits on the $(q_0,\varepsilon$) plane are obtained  and a cosmic acceleration is allowed with a high confidence level (filled region) even if there is some cosmic achromatic absorption mechanism. In this case we obtain: $q_0=-0.62 \pm 0.1 \pm 0.15 \pm 0.27$ for 68.3\%, 95.4\% and 99.99\% c.l.. In Fig.(4b) we plot the likelihood of $q_0$ by marginalizing on the $\varepsilon$ parameter. The horizontal lines correspond to 1$\sigma$ c.l. (68.3\%) and 2$\sigma$ c.l. (95.4\%). Now, as one may see, the intervals of $q_0$ by using SNe Ia, GRBs and $T_{CMB}(z)$ data (black line) are fully within those supporting the cosmic acceleration: $q_0<0$. By considering only SNe Ia and GRB observations we obtain that $q_0 > 0$ is rule out at 2$\sigma$ c.l..

\section{Conclusions}

As it is well known, the  dimming of the distant SNe Ia  has been interpreted as a consequence of the present accelerated stage of evolution of the Universe. However, such an interpretation depends on  a assumption of a transparent universe . In fact, an absorption source leads to a reduction of the photon number from a luminosity source by a factor of $e^{-\tau(z)}$ and, consequently, an increase of its inferred luminosity distance. We have considered $\tau(z)=2\ln(1+z)^{\varepsilon}$, where $\tau(z)$ denotes the opacity  between an observer at $z=0$ and a source at $z$. 

In this work, we have reviewed the well-known result in which only SNe Ia data in a flat $\Lambda$CDM model allows the Einstein-de Sitter model (without acceleration) if some opacity source exists (see Fig.2a, black solid line). Moreover, by considering a flat kinematic approach, we have also verified that a positive deceleration parameter ($q_0 >0$) is allowed within $95.4\%$ (see Fig. 3a, black solid line). However, by considering a joint analysis of SNe Ia along with gamma-ray burst data, which is also affected if some cosmic opacity source is present, we have shown that observations rule out a decelerated universe ($q_0>0$) at $95.4\%$ c.l. even in the presence of cosmic opacity (see Fig.4b).  We have also obtained that $\Omega_{\Lambda} \geq 0.33$ is favored by these data in flat $\Lambda$CDM model  (see Fig.4a). In these figures we have marginalized on the $\varepsilon$ parameter.

 On the other hand, under the assumption of an  absorption probability that is independent of photon wavelength, we have added $T_{CMB}(z)$ measurements to the analyses via a deformed temperature evolution law, and  the cosmic acceleration evidence is reinforced (see filled contours in Figs. 2b and 3b). These joint analyses were performed by using the following deformed equations related to the cosmic distance duality relation and the evolution law of CMB: $D_LD_A^{-1}=(1+z)^{2+\varepsilon}$ and $T_{CMB}(z)=T_0(1+z)^{1-\beta}$, where $\beta=-\frac{2}{3}\varepsilon$ if one considers an absorption probability that is independent of photon wavelength. It is important to stress that such an assumption is well verified for current analyses by using observational data at different wavelengths (see Table I). Finally, it is notable that three different types of observations that are affected by cosmic opacity provide an accelerated universe when analyzed together in a framework without the assumption of cosmic transparency. 
 
%From a theoretical point of view the joint analysis shows that some source of acceleration (as a %cosmological constant or some mechanism that mimic it) must be present even if some opacity are %present.

%%%%%%%%%%%%%%%%%%%%%%%%%%%%%%%%%%%%%%%%%%%%%%%%%%%%%%%%%%%%%%%%%%%%%%%
\begin{acknowledgements}
RFLH acknowledges financial support from  CNPq  (No. 303734/2014-0). SHP  acknowledges financial support from CNPq - Conselho Nacional de Desenvolvimento Cient\'ifico e Tecnol\'ogico, Brazilian research agency, for financial support, grants number 304297/2015-1 and 400924/2016-1. One of the author (DJ) acknowledges the support provided by IUCAA (Pune) under the visiting Associateship Programme where part of work has been done. 
\end{acknowledgements}
%%%%%%%%%%%%%%%%%%%%%%%%%%%%%%%%%%%%%%%%%%%%%%%%%%%%%

%%%%%%%%%%%%%%%%%%%%%%%%%%%%%%%%%%%%%%%%%%%%%%%%%%%%%%%%%%%%%%%%%%%%%%%%%%

\end{document}